# Predicting Propensity to Vote with Machine Learning

Rebecca D. Pollard[a]*, Sara M. Pollard[b] and Scott E. Streit[c]

[a]*Maret School, Washington DC, USA;*

[b]*Bullis School, Potomac, MD, USA*

[c]*Private Identity LLC, Woodbine, MD, USA*

*corresponding author Rebecca Pollard bpollard@maret.org

Rebecca Pollard is a first-year high school student at Maret School in Washington, DC.

Sara Pollard is in her last year of high school at Bullis School in Potomac, MD.

Scott Streit is Chief Technology Officer of Private Identity LLC, Chair of Biometrics for IEEE and Chair of IEEE Standard 2410 (BOPS). Earlier in his career he served as Chief Scientist for a US Government agency and Principal Investigator and professor at Villanova University.

# Predicting Propensity to Vote with Machine Learning

We demonstrate that machine learning enables the capability to infer an individual's propensity to vote from their past actions and attributes. This is useful for microtargeting voter outreach, voter education and get-out-the-vote (GOTV) campaigns. Political scientists developed increasingly sophisticated techniques for estimating election outcomes since the late 1940s. Two prior studies similarly used machine learning to predict individual future voting behavior. We built a machine learning environment using TensorFlow, obtained voting data from 2004 to 2018, and then ran three experiments. We show positive results with a Matthews correlation coefficient of 0.39.

**Introduction**

Increasingly sophisticated models of the voting process and techniques for estimating election outcomes were present since the 1940s [1]. Inaccuracies and methodological problems existed in these surveys and included sampling error, response validity, timing of the field work, estimation of the likely electorate, and allocation of the undecided in the distribution of candidate preference [2]. Real-time polling resolves many of these issues but is expensive and rarely done [3].

Since 2015, studies noted the increased prevalence of machine learning (ML) in political science research [4]. These researchers leveraged ML to improve the accuracy of outcomes, refine measurements of complex processes, address non-linearities in data and introduce new kinds of data [5].

Two ML approaches, supervised and unsupervised, are relevant in political science research. Unsupervised ML, as used below in this paper, finds patterns in data by modeling density estimations using only independent variables and without prior knowledge of the dependent variable. Unsupervised ML enables us to find relevant patterns and causal mechanisms[6].

Two prior efforts applied ML techniques to predict an individual voter's propensity to vote. Pew Research's effort in 2016 used decision trees and machine learning to predict each voter's propensity to vote as a method to validate election survey answers [7]. The most impressive effort, however, was by Tynan Challenor, at the time an undergraduate student at Stanford University, who trained three models using census data and reported spectacular results [8].

**Methods**

First, we set up a host and target computer system to perform ML training and analysis. The host prepares and trains the data. The target runs the predictions. In our case, the host and target were the same machine. Specifically, the host machine had 6 gigabytes of GPU, 64 gigabytes of RAM, 2 terabytes of solid state storage and an Ubuntu 20.04 operating system with MySQL, Cuda, Cudnn and Tensorflow-gpu.

Next, we downloaded the IPUMS-ASA U.S. Voting Behaviors data from the American Statistical Society (Fall Data Challenge, 2020) [9]. We then built a database schema to store the IPUMS-ASA U.S. Voting Behaviors training data and loaded the data into MySQL using the following steps. We created the database, tables and a user with normal operating system privileges. This keeps the concept of least privilege. After that, we wrote a short bash script that calls a Python program that loads the data from the csv file to MySQL.

Next, we trained a TensorFlow model capable of training and predicting using the data previously loaded into MySQL. We divide the data into training and validation sets. The training data was from 2004 to 2014 or 2004 to 2016. The validation set was either 2016 and 2018 or only 2018 data.

Finally, we wrote a Python program to train and evaluate a model. This program read from the voter table, trained using that data, and evaluated using the validation set. We used the validation set to check our accuracy of the trained model. We evaluated results using the resulting log files. We changed the data assumptions and ran the experiment again three separate times.

**Results**

The first experiment we ran with no changes made to the downloaded data. We trained the model using data from 2004 to 2016 and evaluated the model using data from 2018. Table 1 shows the log file with results from Experiment 1. Table 2 shows the confusion matrix created from the evaluation data. Table 3 shows experimental results computed using the confusion matrix.

For the second experiment, we added one column of derived data to the data set to indicate presidential election years. We trained the model using data from 2004 to 2014 and evaluated the model using data from 2016 and 2018. The results of this experiment were similar from Experiment 1 so we do not reference a confusion matrix here. Table 4 shows the log file with results from Experiment 2.

For the third experiment, we added 5 additional columns (race, marital status, Hispanic origin, educational level, and whether the subject graduated from high school) to the data set from the second experiment in an attempt to further generalize the data (see Table 5). We trained the model using data from 2004 to 2014 and evaluated the model using data from 2016 and 2018. Table 6 shows the log file with results from

Experiment 3. Table 7 shows the confusion matrix created from the evaluation data. Table 8 shows experimental results computed using the confusion matrix.

In the third experiment, we also eliminated the following variables as distractors : EDHGCGED, VOWHYNOT, VOTEHOW, VOTEWHEN, VOTEREGHOW, VOREG AND VOSUPPWT.

**Discussion**

By inferring an individual's propensity to vote from their past actions, political campaigns and governments can micro target voter outreach, education and get-out-the-vote (GOVT) campaigns. Against this background, in this article we attempt to accurately predict voter behaviour using unsupervised machine learning and simple voter attributes. With our setup, we aim to predict the likelihood of each registered voter to vote.

On a general level, our predictions of voting show modest levels of success. Here, we rely on the Matthews correlation coefficient (MCC), a unified statistical rate that represents the quality of a binary prediction [10]. In particular, Experiment 1 had an MCC of 0.39, indicating a positive correlation. In experiments two and three, we added contrived data in an attempt to improve accuracy. Here, the MCC shows that our attempts to improve the results were unsuccessful.

Overall, only the Challenor [8] study attempts a similar prediction. We compare our results to this study and find that Challenor achieved superior results using a Support Vector Machine (SVM). We calculated the MCC using the confusion matrix shown in Challenor's report and found MCC = 0.74. In the future, we will attempt to increase the correlation by expanding our data sources.

```
train_2020-11-05-1539_SPLIT_2016_NO_PRES_69_percent.log
INCLUDE_PRESIDENT_ELECTION_ATTRIBUTE: False
SPLIT_YEAR: 2016
INFO:tensorflow:Evaluation [10/100]
INFO:tensorflow:Finished evaluation at 2020-11-05-16:00:09
INFO:tensorflow:Saving dict for global step 1500: acc = 0.68635815, global_step = 1500, loss = 0.7592631
```

Table 1.  Experiment 1 – Results

|  | Predicted: NO | Predicted: YES |  |
|---|---|---|---|
| **Actual: NO** | 15052 | 9615 | **24667** |
| **Actual: YES** | 12563 | 35774 | **48337** |
|  | **27615** | **45389** |  |

Table 2.  Experiment 1 - Confusion Matrix

| Measure | Value | Derivations |
|---|---|---|
| Sensitivity | 0.5451 | $TPR = TP / (TP + FN)$ |
| Specificity | 0.7882 | $SPC = TN / (FP + TN)$ |
| Precision | 0.6102 | $PPV = TP / (TP + FP)$ |
| Negative Predictive Value | 0.7401 | $NPV = TN / (TN + FN)$ |
| False Positive Rate | 0.2118 | $FPR = FP / (FP + TN)$ |
| False Discovery Rate | 0.3898 | $FDR = FP / (FP + TP)$ |
| False Negative Rate | 0.4549 | $FNR = FN / (FN + TP)$ |
| Accuracy | 0.6962 | $ACC = (TP + TN) / (P + N)$ |
| F1 Score | 0.5758 | $F1 = 2TP / (2TP + FP + FN)$ |
| Matthews Correlation Coefficient | 0.3417 | $TP*TN - FP*FN / sqrt((TP+FP)*(TP+FN)*(TN+FP)*(TN+FN))$ |

Table 3.  Experiment 1 – Analysis of Confusion Matrix

train_2020-11-05-1632_INCLUDE_PRES_SPLIT_2016.log
INCLUDE_PRESIDENT_ELECTION_ATTRIBUTE: True
SPLIT_YEAR: 2016
INFO:tensorflow:Finished evaluation at 2020-11-05-16:39:21
INFO:tensorflow:Saving dict for global step 500: **acc = 0.6743967**, global_step = 500, loss = 0.7775075

Table 4. Experiment 2 – Results

| Variable | Consolidated Categories |
|---|---|
| RACE | White, Black, Asian and other |
| MARST | Yes or No |
| HISPAN | Yes or No |
| EDUC99 | High School or less, Beyond HS, Graduated |
| EDDIPGED | GED or High school graduate |

Table 5. Columns consolidated in an attempt to generalize the training data

train_2020-11-06-1133_WITH_CODES_CHANGES.log
INCLUDE_PRESIDENT_ELECTION_ATTRIBUTE: False
SPLIT_YEAR: 2016
INFO:tensorflow:Evaluation [10/100]
INFO:tensorflow:Finished evaluation at 2020-11-06-11:55:10
INFO:tensorflow:Saving dict for global step 1500: **acc = 0.67716455**, global_step = 1500, loss = 0.76777804

Table 6. Experiment 3 – Results

|  | Predicted: NO | Predicted: YES |  |
|---|---|---|---|
| **Actual: NO** | 19476 | 5191 | **24667** |
| **Actual: YES** | 21636 | 26701 | **48337** |
|  | **41112** | **31892** |  |

Table 7. Experiment 3 – Confusion Matrix

| Measure | Value | Derivations |
|---|---|---|
| Sensitivity | 0.4737 | TPR = TP / (TP + FN) |
| Specificity | 0.8372 | SPC = TN / (FP + TN) |
| Precision | 0.7896 | PPV = TP / (TP + FP) |
| Negative Predictive Value | 0.5524 | NPV = TN / (TN + FN) |
| False Positive Rate | 0.1628 | FPR = FP / (FP + TN) |
| False Discovery Rate | 0.2104 | FDR = FP / (FP + TP) |
| False Negative Rate | 0.5263 | FNR = FN / (FN + TP) |
| Accuracy | 0.6325 | ACC = (TP + TN) / (P + N) |
| F1 Score | 0.5922 | F1 = 2TP / (2TP + FP + FN) |
| Matthews Correlation Coefficient | 0.3261 | TP*TN - FP*FN / sqrt((TP+FP)*(TP+FN)*(TN+FP)*(TN+FN)) |

Table 8. Experiment 3 – Analysis of Confusion Matrix